\documentclass[conference]{IEEEtran}
\IEEEoverridecommandlockouts
\usepackage{cite}
\usepackage{amsmath,amssymb,amsfonts,bm,gensymb}
\usepackage{algorithmic}
\usepackage{graphicx}
\usepackage{textcomp}
\usepackage{xcolor}
\def\BibTeX{{\rm B\kern-.05em{\sc i\kern-.025em b}\kern-.08em
    T\kern-.1667em\lower.7ex\hbox{E}\kern-.125emX}}

\begin{document}

\title{Data-Driven Incident Detection in Power Distribution Systems\\
\thanks{SANDXXXXX2020} 
}
\author{
\IEEEauthorblockN{Nayara Aguiar, Vijay Gupta}
\IEEEauthorblockA{\textit{Department of Electrical Engineering} \\
\textit{University of Notre Dame}\\
Notre Dame, IN, USA \\
\{ngomesde,vgupta2\}@nd.edu}
\and
\IEEEauthorblockN{Rodrigo D. Trevizan, Babu R. Chalamala}
\IEEEauthorblockA{\textit{Energy Storage Technology \& Systems} \\
\textit{Sandia National Laboratories}\\
Albuquerque, NM, USA \\
\{rdtrevi,bchalam\}@sandia.gov}
\and
\IEEEauthorblockN{Raymond H. Byrne}
\IEEEauthorblockA{\textit{Electric Power Systems Research} \\
\textit{Sandia National Laboratories}\\
Albuquerque, NM, USA \\
rhbyrne@sandia.gov}
}

\maketitle

\begin{abstract}
In a power distribution network with energy storage systems (ESS) and advanced controls, traditional monitoring and protection schemes are not well suited for detecting anomalies such as malfunction of controllable devices. In this work, we propose a data-driven technique for the detection of incidents relevant to the operation of ESS in distribution grids. This approach leverages the causal relationship observed among sensor data streams, and does not require prior knowledge of the system model or parameters. Our methodology includes a data augmentation step which allows for the detection of incidents even when sensing is scarce. The effectiveness of our technique is illustrated through case studies which consider active power dispatch and reactive power control of ESS.      
\end{abstract}

\begin{IEEEkeywords}
Energy storage systems, incident detection, power distribution systems.
\end{IEEEkeywords}

\section{Introduction}
Power Distribution Systems (PDS) have been historically designed to transport power from the bulk transmission system to end consumers of electricity. However, the recent increase in the adoption of Distributed Energy Resources (DERs) has started to shift this paradigm, with consumers and utilities leveraging these resources at increasing rates. This high penetration of DERs in PDS will create multiple technical challenges associated with new operating characteristics such as bidirectional power flows and voltage fluctuations due to the volatility of renewable power generation. Specifically, in a distribution system with Energy Storage Systems (ESS), renewable generation and advanced controls, traditional monitoring and protection schemes are not well suited for detecting faults, changes in topology, and malfunction of controllable devices. Furthermore, the methods that rely on accurate measurements and knowledge of parameters currently used in transmission systems, such as traditional power system state estimation, are inadequate for monitoring in PDS. Thus, it is essential to develop new methodologies to detect abnormities in PDS due to critical events (e.g., natural disasters, physical and cyber-attacks) and to mitigate the consequences of these abnormities.  

Classical approaches for incident detection in the transmission grid rely on some knowledge of the system model. For example, the integer programming approach in \cite{tate}, compressive sensing based approach in \cite{zhu}, quickest change detection method in \cite{banerjee}, and the Gauss-Markov graphical model in \cite{he} all rely on some knowledge of grid parameters and of the assets connected to the grid. As compared to works that study the transmission system, the body of literature that considers incident detection in the distribution grid is not as large. Despite the differences between both grids, such as the fact that distribution grid models are little known and sensing is scarce, studies that aim to detect events in the distribution grid started by also assuming knowledge of the network topology, such as in graph-based techniques and methods based on traveling waves (see \cite{bahmanyar} for a review of methods for fault and outage area detection in distribution grids). With the introduction of PMUs, data-driven techniques started receiving more attention, as in the case with SVD-based approaches \cite{lim}. However, most of the focus has still been on the transmission grid, where sensing is more abundant and uniform. When it comes to distribution grids, an increase in the use of micro PMUs and AMIs has been observed, but the existence of heterogeneous sources of data has also posed challenges.

Our proposed approach draws on Koopman operator theory, which accounts for the causal relationship among multiple sensor data streams without prior knowledge of the dynamic model. The major feature of the proposed approach lies on the detection of events which produce sharp changes in the causal map of the system dynamics, while being robust to small variations such as measurement noise and load fluctuation. Koopman operators have been explored in the power systems literature for the identification of system dynamics, stability assessment, and other topics \cite{sinha,sinha2,susuki,susuki2,korda}. In this work, we propose an algorithmic approach that provides evidence of the effectiveness of this operator in incident detection tasks relevant to distribution systems. We also tackle the issue of sensor availability by proposing a way to augment the dataset through a transformation that computes the relationship between data streams. Our case studies explore the detection of incidents in ESS operation that relates both to active power dispatch and reactive power control, while accounting for the presence of measurement noise and load fluctuations.

The remainder of this paper is organized as follows. We introduce the Koopman operator theory in Section~\ref{sec:theory}, present the proposed algorithm for incident detection in Section~\ref{sec:algorithm}, and detail the steps in the methodology for incident detection in Section~\ref{sec:methodology}. Section~\ref{sec:cases} presents two representative case studies which exemplify the use of the proposed algorithm, and concluding remarks are presented in Section~\ref{sec:conclusion}.

\section{Koopman Operator Theory Preliminaries}\label{sec:theory}

Consider a nonlinear system with states $x_t \in \mathbb{R}^n$ whose dynamics can be characterized by
\begin{equation}
    x_{t+1} = \bm{F}(x_t).
\end{equation}
The Koopman operator $\bm{K}$ is a linear infinite-dimensional operator that acts on the space of observables $g(x): \mathbb{R}^n \to \mathbb{R}$, i.e. the space of scalar-valued functions of the states of the system, as follows:
\begin{equation}
    \bm{K}g(x) = g(x) \circ \bm{F}.
\end{equation}

The map $\bm{F}$ can be described as a linear combination of the eigenfunctions of the linear infinite-dimensional operator $\bm{K}$. Then, the evolution of the state becomes a linear combination of independent dynamics along each eigenfunction. Assume that the dominant eigenfunctions are roughly in the span of the $D$ dictionary functions
\begin{equation}
    \bm{\Psi} = (\Psi_1,...,\Psi_D)^T.
\end{equation}
Then, the dynamics of $\bm{F}$ becomes roughly linear if lifted to a space in which the functions in $\bm{\Psi}$ are roughly taken as coordinates. This linear description is characterized by $\mathcal{K}\in \mathbb{R}^D \times \mathbb{R}^D$, which is a finite-dimensional approximation of the Koopman operator. An observable $g(x)$ in the span of dictionary functions identified by weights $b$ is given by $g(x) = \bm{\Psi}(x)^Tb$. Under the action of the Koopman operator, this observable can be approximated as
\begin{equation}
    g(\bm{F}(x)) \approx \bm{\Psi}(x)^T\mathcal{K}b.
\end{equation}

Techniques such as the Extended Dynamic Mode Decomposition (EDMD) algorithm proposed in \cite{williams} provide a way to calculate an approximation of the Koopman operator by minimizing $||\bm{\Psi}(\bm{F}(x))^Tb-\bm{\Psi}(x)^T\mathcal{K}b||^2_2$ for an arbitrary $b$. Let $(x_1,x_2),...,(x_M,x_{M+1})$ be a time-series data. The minimization problem of interest is equivalent to
\begin{equation}
    \underset{\mathcal{K}\in \mathbb{R}^D \times \mathbb{R}^D}{\min} ||\bm{A}-\bm{G}\mathcal{K}||^2_F,
\end{equation}
where $||.||_F$ is the Frobenius norm, and $\bm{A}$ and $\bm{G}$ are defined~as
\begin{equation}
    \bm{A} := \frac{1}{M}\sum_{j=1}^M \bm{\Psi}(x_j)\bm{\Psi}(x_{j+1})^T,~ \bm{G} := \frac{1}{M}\sum_{j=1}^M \bm{\Psi}(x_j)\bm{\Psi}(x_j)^T. 
\end{equation}

\section{Algorithm for Incident Detection}\label{sec:algorithm}
In this work, we leverage a sparsity-promoting variant of the EDMD algorithm \cite{quade} for the detection of incidents in the distribution grid, given by
\begin{equation}\label{eq:ksparse}
    \underset{\mathcal{K}\in \mathbb{R}^D \times \mathbb{R}^D}{\min} ||\bm{A}-\bm{G}\mathcal{K}||^2_F + \lambda ||\text{vec}(\mathcal{K})||_1.
\end{equation}

For our algorithm, we consider the following notation. For time-series data $x$ and $t_1\leq t_2$, define
\begin{equation}
    x[t_1:t_2] := (x_{t_1}x_{t_1+1},...,x_{t_2}).
\end{equation}
At time $t$, let $\bm{A}[t-T:t]$ and $\bm{G}[t-T:t]$ be computed with $T+1$ observations and their one-step propagation in the observed data series $x[t-T:t]$.

Our goal is to maintain a sparse representation of the dynamic system, and detect incidents by detecting changes in this sparsity pattern. For that, we build on \eqref{eq:ksparse} to define our algorithm for incident detection as follows:
\begin{itemize}
    \item At time $t=T+1$, compute $\mathcal{K}_{t}$ by solving
    \begin{equation}
        \underset{\mathcal{K}_t\in \mathbb{R}^D \times \mathbb{R}^D}{\min} ||\bm{A}[1:T+1]-\bm{G}[1:T+1]\mathcal{K}_t||^2_F + \alpha ||\text{vec}(\mathcal{K}_t)||_1.
    \end{equation}
    \item From $t=T+2$ onward, compute $\mathcal{K}_{t}$ by solving
    \begin{equation}
    \begin{split}
        \underset{\mathcal{K}_t\in \mathbb{R}^D \times \mathbb{R}^D}{\min} &||\bm{A}[t-T:t]-\bm{G}[t-T:t]\mathcal{K}_t||^2_F \\ &+\alpha ||\text{vec}(\mathcal{K}_t)||_1+\beta ||\mathcal{K}_{t-1}-\mathcal{K}_t||_F.
    \end{split}
    \end{equation}
    \item If the sparsity pattern of $\mathcal{K}_{t}$ differs significantly from that of $\mathcal{K}_{t-1}$, flag the occurrence of an incident that has altered the causal map between a state and its one-step propagation.
\end{itemize}

The sparsity pattern of each $\mathcal{K}$ encodes the causal relationship among observables, and the level of sparsity is adjusted by the choice of the hyper-parameter $\alpha$. Small values of $\alpha$ define more complete networks, while large values promote sparsity. The hyper-parameter $\beta$ is related to the smoothness of transitions. Thus, larger values of $\beta$ tend to stabilize $\mathcal{K}$ along a trajectory, making its less prone to change due to smaller variations which do not lead to major structural changes.

\section{Methodology for Incident Detection}\label{sec:methodology}
\begin{figure}[h]
\centering
\includegraphics[width=0.5\textwidth]{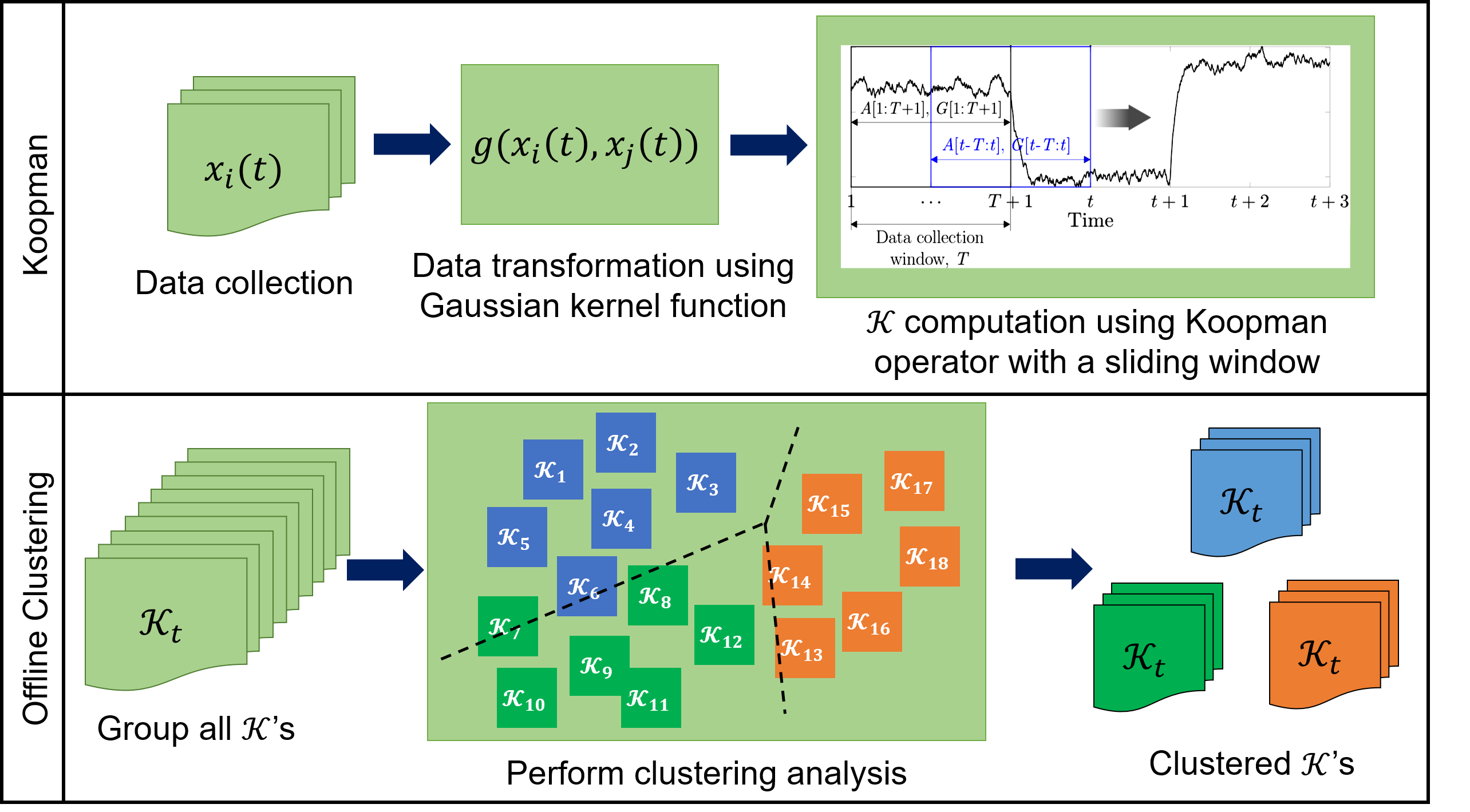}
\caption{Two-part methodology developed for the detection of incidents in distribution grids.}
\label{fig:method}
\end{figure}

Our methodology can be divided into two distinct parts, as illustrated in Fig.~\ref{fig:method}. The first part consists of a data collection step, followed by a data transformation. The transformed data is then used to compute the $\mathcal{K}$ matrices using the sliding window approach introduced in Section~\ref{sec:algorithm}. The second part is a post-processing clustering task which performs a clustering analysis with the goal of grouping the $\mathcal{K}$ matrices calculated into clusters that are uniquely identified by one of the scenarios simulated in the data collection step. Each part of this approach is discussed in more detail in the next sections.  

\subsection{Part 1: Koopman Operator}
\subsubsection{Data Collection}
We begin by selecting suitable simulation scenarios from which our data will be gathered. Since our focus is on incidents related to the operation of ESS in the distribution grid, the following two broad types of events were considered:
\begin{enumerate}
    \item Changes in ESS charging/discharging rate: In distribution systems, ESS can be used for peak-shaving operations, absorption of excess local renewable production, and other tasks which involve a coordinated charging/discharging schedule. We aim to analyze whether changes in the charging/discharging rates of ESS can be identified with our proposed framework. Detecting such changes would allow us to recognize, for example, when a battery fleet is no longer following the command issued for their power supply/absorption operation.
    
    \item Changes in ESS controller parameters: Voltage regulation is an important task in power distribution systems, especially with the increasing adoption of DERs. By considering a scenario in which batteries offer voltage regulation services, we model Volt/VAR curves to control the reactive power output of these devices. The goal is to detect changes in controller parameters, which could have been caused by attacks that aim to destabilize the grid.
\end{enumerate}

We assume the data collected comes from PMUs, and thus we use voltage magnitude and phase measurements. The data was collected with a 0.25s sampling time. Because sensor measurements are typically noisy, Gaussian noise was added to the voltage magnitude ($\pm 0.01\%$) and voltage angle ($\pm 0.01\degree$) measurements. Further, considering that sensing capability would be limited in the distribution grid, we assumed that only nodes with an ESS had PMUs. These scenarios were simulated using OpenDSS.

\subsubsection{Data Transformation}
We hypothesize that the relationship between different time-series data carries more information than individual data streams. In \cite{fujii}, this dependence among observations is explored in a data-driven spectral analysis using the Koopman operator with the objective of understanding complex biological network dynamics. For this purpose, using a ballgame as an example, the authors transformed the data using the Gaussian kernel
\begin{equation}\label{eq:gaussian}
    g(x_i,x_j) = \exp\left(-\frac{||x_i-x_j||^2_2}{2\sigma}\right),
\end{equation}
where each $x_i$ is a stream of raw data, $||.||_2$ is the Euclidean norm, and $\sigma$ is an adjustment parameter.

We apply \eqref{eq:gaussian} to the distribution grid data to be used in our analysis. This nonlinear transformation lifts the data into a higher dimensional space by giving a measure of similarity between states, thus augmenting our dataset. The transformed data is then used to compute a sequence of $\mathcal{K}$ matrices through a sliding window.

\subsubsection{Approximate Koopman Operator}
In this step, the finite-dimensional approximation of the Koopman operator for the transformed data was performed numerically following the steps detailed in Section~\ref{sec:algorithm}. Significant changes in $\mathcal{K}$ indicates altered causality in states (i.e., occurrence of incidents). Radial basis functions of the form
\begin{equation}\label{eq:rbf}
    \Psi_i(x) = ||x-c_i||^2\log_e(||x-c_i||),
\end{equation}
where $c_i$ are unique center points, have shown to be effective for our application. We face the following trade-off when choosing the amount of dictionary functions to be used: too few dictionary functions $\Psi_i(x)$ may lead to poorer distinguishability of the sparsity pattern of $\mathcal{K}$'s from different incidents; increasing the amount of functions, however, also increases computational time.

\subsection{Part 2: Offline Clustering}
The previous step was aimed at generating data for different case studies, transform the data collected, and compute the approximate Koopman operator for multiple time windows. The offline clustering task performs a clustering analysis in the $\mathcal{K}$ matrices calculated. Since the sparsity pattern of these matrices is expected to be similar if the system dynamics remains the same, this analysis is expected to cluster together the $\mathcal{K}$'s coming from the same simulated scenario. For example, the matrices calculated from a scenario in which a battery fleet is discharging at 
25\% rate should be clustered together, while $\mathcal{K}$'s from a scenario where these batteries are discharging at 100\% should be together but in a different cluster.

We used the k-means clustering algorithm for this analysis. The idea behind k-means is that observations within a cluster are close to each other, while observations in different clusters are far apart. This method takes as inputs a distance measure, the number of clusters, the data to be clustered, and a random seed that is used to initialize the algorithm. For our application, the correlation distance was shown to perform better than other metrics. Further, prior to the clustering analysis, the $\mathcal{K}$ matrices were transformed into binary matrices, i.e. values below a certain threshold were set to zero and all others were set to one. The purpose behind this transformation is that we are only concerned with the sparsity pattern of these matrices, and not with the actual values of their elements. This step also avoids numerical issues that may arise when working with small numbers. Since the approximate Koopman operator is a representation of our original data in a high-dimensional space, it is common to see elements of $\mathcal{K}$ which have a really small order of magnitude. The effectiveness of our results is evaluated based on the misclassification rate achieved in this clustering analysis.  

\section{Case Studies}\label{sec:cases}
We considered the IEEE 8500-node test feeder with 7 battery energy storage systems (BESS) added, implemented using OpenDSS. This feeder is an unbalanced radial network, which are typical characteristics of distribution grids. These BESS can take real power dispatch commands and perform Volt/VAR control. The data was collected with a 0.25s sampling time, and Gaussian noise was added to the voltage magnitude ($\pm 0.01\%$) and voltage angle ($\pm 0.01\degree$) measurements. Besides noise, fluctuations in real and reactive load were considered in the feeder model. Our aim is to evaluate the robustness of our algorithm when these variations are accounted for. 

As previously discussed, changes in the sparsity pattern of the $\mathcal{K}$ matrices indicated the occurrence of an incident. To confirm the $\mathcal{K}$ matrices had unique sparsity patterns for each scenario in which the causality of the model was maintained, an offline analysis was performed using k-means clustering to cluster these matrices. For both case studies, the centers for the radial basis functions $c_i$ in \eqref{eq:rbf} were chosen in the computation of the first $\mathcal{K}$ matrix, and then kept constant throughout the experiment. The center points were selected using a randomly generated perturbation, so that they were of the same order of magnitude as the data itself. Further, we used $D=400$ dictionary functions, a time window of $T=100s$, and a new $\mathcal{K}$ was estimated every $25s$. We also assumed voltage magnitude and phase measurements from PMUs were available at the nodes of 3 of these BESS.

\subsection{Detecting changes in discharging rates}

\begin{figure}[h]
\centering
\includegraphics[width=0.5\textwidth]{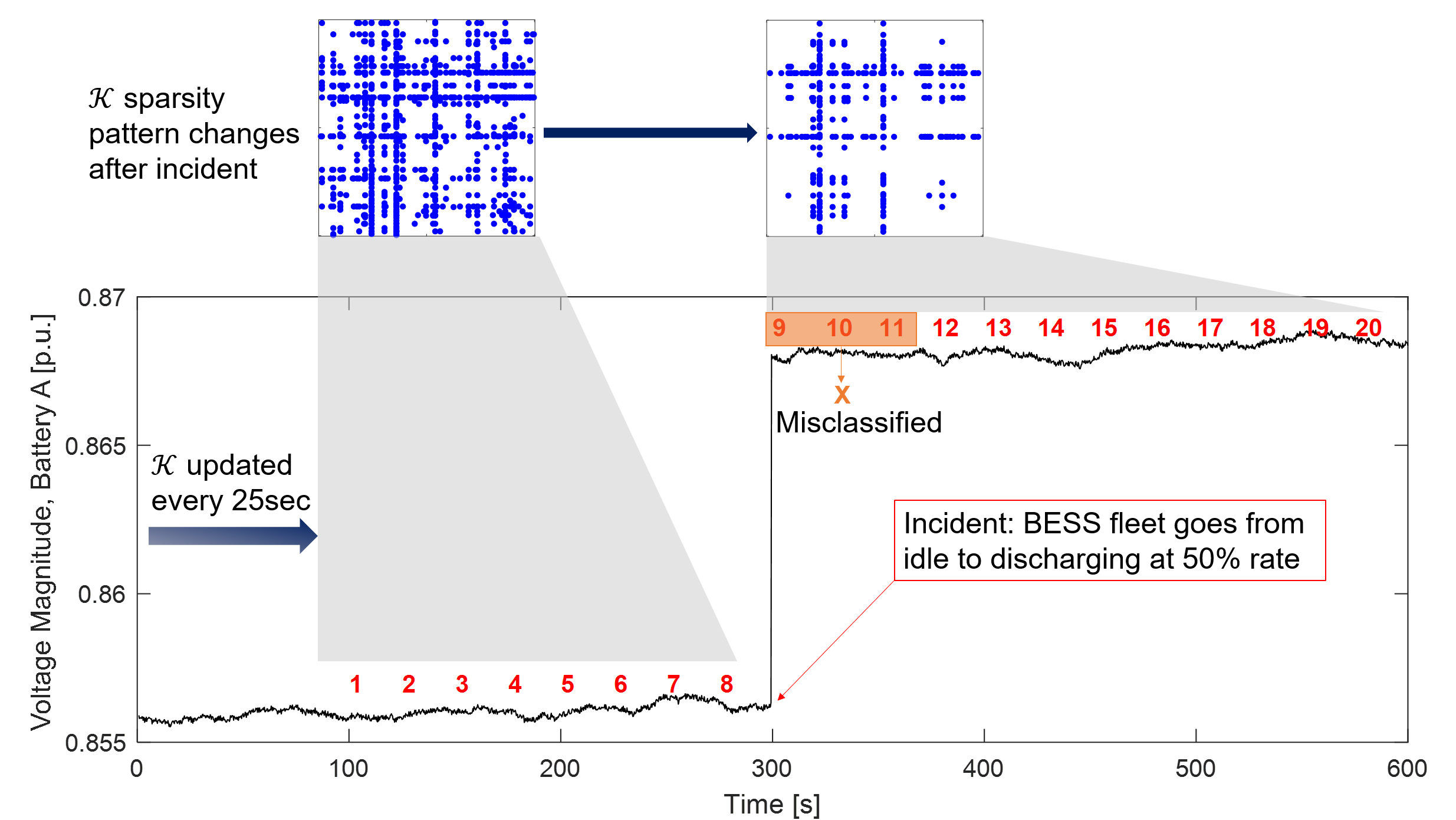}
\caption{Results for case study which aims to detect changes in the rate at which the battery fleet supplies active power to the grid.}
\label{fig:case1}
\end{figure}

We first consider a scenario in which the battery fleet is initially idle. After 5min, all the batteries start discharging at 50\% of their discharging rate, maintaining this rate for 5min. Fig.~\ref{fig:case1} shows the results for this case. For the Koopman operator approximation step, we let $\alpha=0.1$ and $\beta=0.4$.

The red numbers in Fig.~\ref{fig:case1} mark the time instants at which a new $\mathcal{K}$ matrix was estimated. The plots on top show a representative sparsity pattern for each of the two situations considered, and we can clearly observe that this pattern changes after the incident happens, allowing us to successfully detect its occurrence. To corroborate that the two patterns are indeed distinct, all the 20 $\mathcal{K}$ matrices computed were subjected to a clustering analysis using k-means. For this case study, we have two different clusters: the first representing our network dynamics when the batteries are idle, and the second related to the case in which the battery fleet discharges at 50\% of their rate. As a result, we found that 3 of these matrices were misclassified, i.e. they were classified to the cluster that does not correspond to the scenario they actually refer to. We note that the misclassified matrices, highlighted in Fig.~\ref{fig:case1}, were computed immediately after the event occurred. Due to the sliding window approach, their computation used data from both the first and the second scenarios, which impacts their sparsity pattern. However, the matrices after those were correctly classified. Thus, we can state that some time may be needed after the event for the new sparsity pattern to stabilize. We remark that the results were robust to measurement noise, as well as load variations, since these fluctuations did not lead our algorithm to erroneously identify the occurrence of nonexistent incidents.   

\subsection{Detecting changes in controller parameters}
\begin{figure}[h]
\centering
\includegraphics[width=0.45\textwidth]{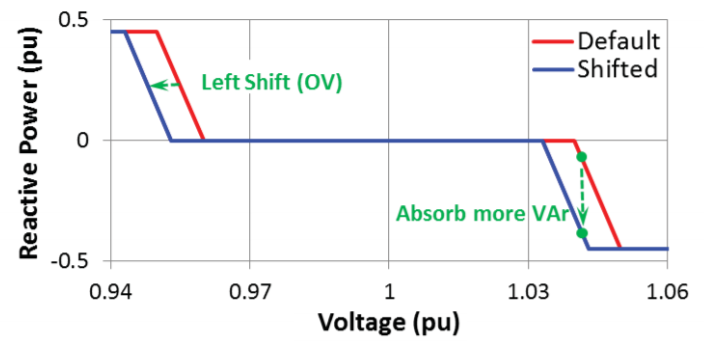}
\caption{Example of VV curve and shifting logic \cite{quiroz}.}
\label{fig:vvcurve}
\end{figure}

In this case study, we considered the battery fleet provides voltage regulation services. For this purpose, the hierarchical Volt/VAR (VV) control strategy proposed in \cite{quiroz} is used to control the reactive power supply/absorption of these devices so that the voltage magnitudes are maintained close to their nominal value. Each battery is assigned a VV curve, which is designed to adjust the reactive power of the device according to measured local voltage levels. An example of this curve is shown in Fig.~\ref{fig:vvcurve} \cite{quiroz}. The VV curve has a deadband corresponding to acceptable voltage levels which do not trigger control action for correction. When voltage levels become high, the battery absorbs reactive power to lower the voltage back to acceptable levels. The opposite happens when voltage levels are low, with the batteries injecting reactive power into the grid. This control layer is local, as each device does not have information about the overall system condition, and only responds to their local condition. 

To overcome this limitation and expand the control strategy to regulate voltage levels throughout the network, a centralized control layer is added. This layer has full information about the voltages over the entire feeder, and is able to dispatch new VV curves to the devices with the goal of regulating the overall voltage in the system, even when the device's local voltage magnitude is within acceptable ranges. Fig.~\ref{fig:vvcurve} shows an example of how the initial VV curve assigned to a battery can be shifted using the proposed centralized strategy. We note that shifting the curve to the left induces an increase in reactive power absorption, which helps decreasing voltage levels more significantly across the network. The reverse holds for right shifts in the curve.

\begin{figure}[h]
\centering
\includegraphics[width=0.5\textwidth]{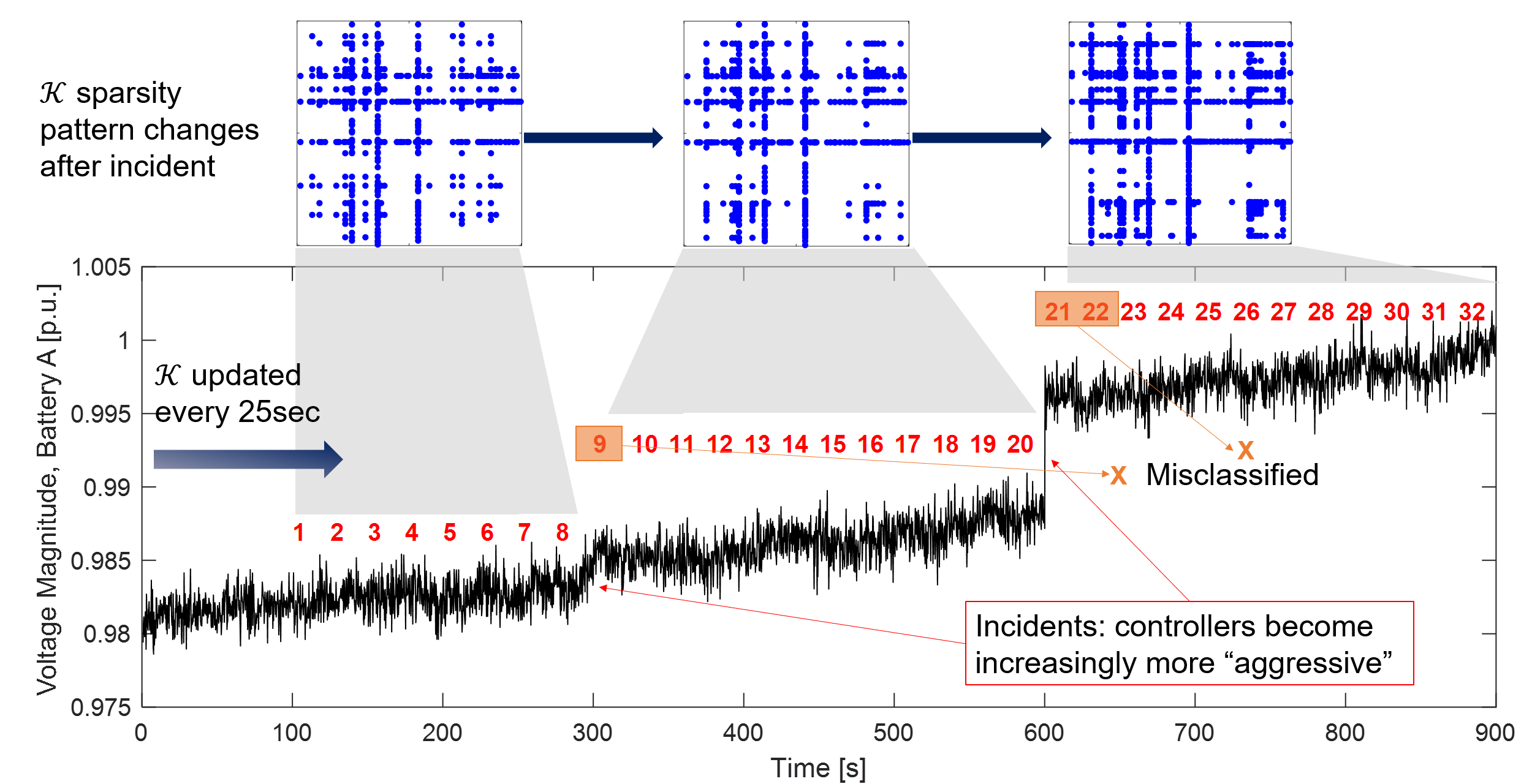}
\vspace{-5mm}
\caption{Results for the detection of changes in the controller parameters of batteries providing voltage regulation through Volt/VAR control.}
\label{fig:case2}
\end{figure}

Using this hierarchical VV control, we modeled two events for this case study, leading to a total amount of three different scenarios. For each event, the controller parameters are changed by adjusting the set-points for the VV curves assigned to the batteries. The dead band of the controller changes from 0.95 to 0.98 and then 0.99 pu, with saturation point shifting right accordingly at $t=200s$ and $t=400$, respectively. The results for this case study are presented in Fig.~\ref{fig:case2}, in which we observe three distinct sparsity patterns for the $\mathcal{K}$ matrices in each scenario. These patterns were achieved by using $\alpha=0.1$ and $\beta=0$ in the Koopman operator approximation. Similarly to the first case study, the differences in the $\mathcal{K}$ matrices were sufficient for us to identify the occurrence of the events considered. Further, in the offline classification task, 3 out of the 32 matrices were misclassified, all of which were computed in time instants immediately following an event. 

\section{Conclusions}\label{sec:conclusion}
We proposed a data-driven method that requires no prior knowledge of the network dynamic model for the detection of incidents in power distribution systems. Our case studies considered the occurrence of changes in BESS operations~that refer both to active power dispatch and reactive power control of these devices. The successful detection of the events modeled allows for proper mitigation strategies to be set forth, if needed. Our methodology introduces a data transformation step which augments the dataset while maintaining meaningful information of the system states. This strategy is particularly useful in situations with restricted sensor availability, which can be common in distribution grids. Further, 
the algorithm proposed was shown to be robust to measurement noise and load fluctuations, and thus such variations do not trigger a~false detection. Future work includes 
designing a more systematic way to choose the hyper-parameters for the Koopman operator approximation, and evaluating our methodology in the presence of data loss or data streams with different~sampling~time.

\section*{Acknowledgment}
This research was funded by the energy storage program at the U.S. Department of Energy under the guidance of Dr. Imre Gyuk. Sandia National Laboratories is a multi-mission laboratory managed and operated by National Technology and Engineering Solutions of Sandia, LLC., a wholly owned subsidiary of Honeywell International, Inc., for the U.S. Department of Energy National Nuclear Security Administration under contract DE-NA-0003525. This paper describes objective technical results and analysis. Any subjective views or opinions that might be expressed in the paper do not necessarily represent the views of the U.S. Department of Energy or the United States Government.
Jamal Al Hourani helped with some numerical experiments. Hyungjin Choi helped with the initial code and explanation of Koopman operator calculations. The work of the first two authors was supported in part by Sandia National Lab under the grant PO 2079716.

\bibliographystyle{IEEEtran}
\bibliography{references.bib}

\end{document}